\documentclass[a4paper,11pt]{article}
\usepackage{jinstpub}
\usepackage{lineno}
\usepackage{import}
\usepackage{float}
\usepackage{algorithm} 
\usepackage{multirow}
\usepackage{algpseudocode}
\usepackage{algorithm}

\title{\boldmath Machine learning evaluation in the Global Event Processor FPGA for the ATLAS trigger upgrade}

\author[a,*]{Zhixing Jiang}
\author[c]{Ben Carlson}
\author[d]{Allison Deiana}
\author[e]{Jeff Eastlack}
\author[a]{Scott Hauck}
\author[b]{Shih-Chieh Hsu}
\author[d]{Rohin Narayan}
\author[d]{Santosh Parajuli}
\author[a]{Dennis Yin}
\author[a]{Bowen Zuo}

\affiliation[a]{Department of Electrical \& Computer Engineering, University of Washington,\\
1410 NE Campus Parkway, Seattle, WA, USA}
\affiliation[b]{Department of Physics, University of Washington,\\
1410 NE Campus Parkway, Seattle, WA, USA}
\affiliation[c]{Department of Physics \& Astronomy, Westmont College and University of Pittsburgh,\\
3941 O'Hara St, Pittsburgh, PA, USA}
\affiliation[d]{Department of Physics, Southern Methodist University,\\
3225 University Blvd., Dallas, TX, USA}
\affiliation[e]{Department of Physics \& Astronomy, Michigan State University,\\
288 Farm Lane, East Lansing, MI, USA}

\emailAdd{zhixij@uw.edu*}

\abstract{The Global Event Processor (GEP) FPGA is an area-constrained, performance-critical element of the Large Hadron Collider's (LHC) ATLAS experiment.  It needs to very quickly determine which small fraction of detected events should be retained for further processing, and which other events will be discarded.  This system involves a large number of individual processing tasks, brought together within the overall Algorithm Processing Platform (APP), to make filtering decisions at an overall latency of no more than 8ms. Currently, such filtering tasks are hand-coded implementations of standard deterministic signal processing tasks.

In this paper we present methods to automatically create machine learning based algorithms for use within the APP framework, and demonstrate several successful such deployments. We leverage existing machine learning to FPGA flows such as \texttt{hls4ml} and \texttt{fwX} to significantly reduce the complexity of algorithm design. These have resulted in implementations of various machine learning algorithms with latencies of $1.2\mu s$ and less than 5\% resource utilization on an Xilinx XCVU9P FPGA. Finally, we implement these algorithms into the GEP system and present their actual performance.

Our work shows the potential of using machine learning in the GEP for high-energy physics applications. This can significantly improve the performance of the trigger system and enable the ATLAS experiment to collect more data and make more discoveries. The architecture and approach presented in this paper can also be applied to other applications that require real-time processing of large volumes of data. }


\begin{document}
\maketitle
\flushbottom

\section{Introduction} 
The ATLAS experiment at the LHC~\cite{Evans:2008zzb} at CERN is undergoing continuous upgrades as part of the High-Luminosity LHC Upgrade~\cite{Apollinari:2015wtw} due to the need to handle an increased data output rate and refine data capture accuracy for the upcoming High-Luminosity LHC upgrade~\cite{ATL-COM-DAQ-2021-093}. The upgrades include a new decision-making module, the Global Trigger subsystem, in the L0 Trigger~\cite{ATLAS-TDR-029}. The L0 trigger is the first-level, hardware-based decision system that selects relevant collision events for further analysis, which will require new and improved hardware and algorithms to increase its performance.

The upcoming Global Trigger subsystem is designed to run advanced algorithms, similar to those typically used for offline data analysis, on detailed data collected from various sub-detectors and processing units in real time. This approach will enhance the quality of detected events and observables, serving as inputs for the advanced decision-making processes handled by the GEP~\cite{firmware_spec}. As the GEP performs many tasks on the same FPGA, the feasible latency for typical individual algorithms is less than $1.2 \mu s$, derived from the $25 ns$ time for each bunch crossing (the time between collisions in the detector) and the number of parallel GEP units receiving data in a round-robin fashion (i.e., 25 ns x 48 GEP units). FPGA resource utilization also must be small enough to incorporate many algorithms, placing practical constraints at the level of a few percent per resource type (LUT, FF, BRAM, DSP).

The GEP, which serves as an FPGA-based framework for an interconnected network of Algorithm Processing Units (APUs), orchestrates the data flow and the processing chain across multiple clock domains to execute the trigger algorithm. Data is pipelined through different APUs within the GEP, with each APU handling individual sub-tasks of the overall trigger. Specialized algorithms are implemented in each APU for data analysis in a pipeline workflow.

The APU emerges as a paradigm of innovation within the ATLAS experiment's data processing systems, demonstrating superior performance over general-purpose processors. Its distinctive advantage lies in utilizing a single FPGA platform to host various algorithms, which streamlines efficiency by obviating the need for cross-platform conversion. With a specialized protocol, the APU facilitates ease of use for designers, enabling seamless integration of multiple APUs where each focuses on a distinct computational challenge. This modular approach, where individual APUs are dedicated to specific tasks and then unified, significantly amplifies the processing capacity of the GEP. Optimized for high-speed processing, the APU surpasses the latency limitations commonly associated with general-purpose processors. Its architecture is intricately designed to manage the complex data flow and algorithmic demands of particle physics experiments, ensuring the delivery of real-time analytics essential for prompt decision-making and dynamic experiment adaptation.

This work is significant because it marks the first time that machine learning tools such as \texttt{hls4ml} and \texttt{fwX} have been used for the ATLAS trigger system. Our paper describes how we deployed these tools into the APU development process, thus simplifying algorithm design and improving APU performance. With the integration of machine learning algorithms into the APU, we have striven toward the theoretical maximum latency of 1.2 microseconds.

The structure of this paper is outlined as follows: Section 2 introduces the APU architecture and communication protocols, details the development process using \texttt{hls4ml} and \texttt{fwX}, and describes the integration of machine learning algorithms into the APU. Section 3 details the results of our experiments and assesses the performance of the GEP-defined algorithms within the APU. The paper concludes with Section 4.

\section{Infrastructure and Methods}
The Algorithm Processing Unit (APU) is an essential part of the Global Event Processor (GEP) system, tasked with the rapid real-time processing and analysis of data from particle detectors. Each APU handles a distinct segment of the total computation, ensuring high-speed data handling that matches the swift transmission rates from the detectors. This section delves into the components essential for managing data transmission and synchronization within the GEP, which are crucial for minimizing data loss and ensuring the accuracy and speed of processing.

\subsection{Integration of the Algorithm}
Machine learning is increasingly prevalent in particle and energy research, particularly in data analysis at the LHC. While not all APU algorithms utilize machine learning, those that do—such as B-tagging algorithms to differentiate jet types from b-quarks using dense or convolutional neural networks, and Quark/Gluon jet tagging—leverage these techniques for enhanced particle identification. However, the FPGA-based firmware design of the APU does not support GPU-based neural network deployments. Instead, tools like \texttt{hls4ml} and \texttt{fwX} are utilized to adapt these neural networks for FPGA implementation. This section discusses the integration of machine learning models into the APU, highlighting the use of an Algorithmic State Machine (ASM) to bridge between the FPGA's firmware architecture and the machine learning models. Both \texttt{hls4ml} and \texttt{fwX} generate Verilog code through Vivado HLS, ensuring consistent structure and protocol across different models and facilitating a standardized approach in applying the ASM.

\begin{algorithm}[t]
\caption{ The ASM for streaming the data input/output to the DNN/BDT}
\label{alg:datatransfer}
\begin{algorithmic}
\State $Param\_Delay \gets n;$
\State $state \gets IDLE;$
\While{$event\_ready$}
\If{$read\_state = IDLE$}
\If{ready}
\State $counter \gets data[0]$ \Comment{the first data contains the index of the last valid data}
\State $read\_state \gets TRANSFER$
\EndIf
\ElsIf{$read\_state = TRANSFER$}
\State $enable\_NN\_in \gets 1;$
\For{$i \gets 0$ to $counter-1$}
\State $data \gets read\_upstream\_BRAM(.addr(i));$
\State $send\_data\_to\_NN(data);$
\EndFor
\State $read\_state \gets IDLE;$
\EndIf
\If{$write\_state = IDLE$}
\If{$NN\_output\_valid$}
\State $write\_state \gets TRANSFER;$
\EndIf
\ElsIf{$write\_state = TRANSFER$}
\State $enable\_apu\_out \gets 1;$
\For{$i \gets 0$ to $counter-1$}
\State $data \gets read\_Dense\_output;$
\State $send\_data\_out(data);$
\EndFor
\State $write\_state \gets END;$
\ElsIf{$write\_state = END$}
\State $send\_data\_out(last\_data\_index);$
\State $event\_done \gets 1;$
\EndIf
\EndWhile
\end{algorithmic}
\end{algorithm}

The ASM's primary function is to manage the protocol differences between the APU's FPGA design, which typically uses an addressable input memory buffer, and the streaming data model inherent to ML models. It ensures seamless data transmission, effectively converting the incoming data into a streaming format compatible with the ML models and formatting the output data for the APU's consumption. This process involves the ASM transitioning through various states – from an initial idle state to active data transfer, and finally to completion – ensuring efficient and accurate data handling.

The uniformity in the ASM design, dictated by the similar structure of the ML models generated by \texttt{hls4ml} and \texttt{fwX}, simplifies the integration process. It allows the APU to handle different types of ML algorithms without requiring significant alterations in the ASM structure or its operational methodology.

The detailed experimental results, which will be discussed in subsequent sections, highlight the effectiveness of integrating these diverse ML models into the APU. These results include comprehensive analyses of resource utilization, latency, and overall performance, demonstrating the practicality and efficiency of this integration approach.

In conclusion, the standardized ASM approach significantly enhances the APU's capability to manage a wide range of computational tasks, thereby bolstering the data processing prowess required for LHC experiments. This integration not only represents a technical achievement but also a crucial step forward in the field of high-energy physics research.

\subsection{Data Transmission and Synchronization}
\begin{figure}[h!]
\centering\includegraphics[width=13cm]{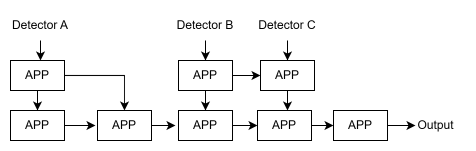}
\caption{\label{GEP_flow}The dataflow of the APUs within the GEP}
\end{figure}

\begin{figure}[h!]
\centering\includegraphics[width=13cm]{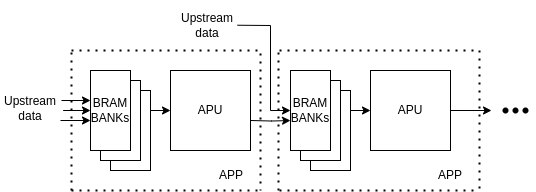}
\caption{\label{dataflow}The communication between two APPs in a detailed view}
\end{figure}

In the GEP, raw input events arrive every $1.2\mu s$, with intervening inputs sent to additional GEP modules. Individual APUs perform portions of the overall computation, with data streaming in a fixed dataflow graph from APP to APP, where an APP is a container for an APU. Parallel paths in this dataflow graph represent different portions of the computation, while parallel execution units for a given step are contained within an individual APU, as demonstrated in figure~\ref{GEP_flow}. BRAM-based buffers are placed between communicating APUs to store the input or output information from each APU and allow parallel operation between the producer and the consumer. As illustrated in figure~\ref{dataflow}, BRAMs are stacked together to form a bank that stores data for multiple events. These data sources can be raw data from the detector or data from an upstream APU. An APU processes one event at a time, receiving data from the upstream BRAMs and storing the resultant data in a downstream BRAM. Fanout in the dataflow graph is supported by parallel copies of the downstream memory buffers.

To address the significant challenge of data synchronization, given the arrival skew of raw data inputs and unsynchronizied clock speeds from the detectors, the APP was developed. The APP serves as a wrapper for each APU and facilitates Clock Domain Crossing (CDC) through its sub-modules.

The APP comprises Synchronization Registers (SR), BRAMs, a Sync controller, and the APU itself. The BRAMs in the APP operate under two clocks: one that writes data from upstream and another that reads data for the APU within the APP. This dual-clock operation enables the transfer of data between different clock speeds. The SR, tasked with determining when data from a particular input source is ready, controls a stack of BRAMs in the APP and governs data storage and retrieval. The Sync controller, which contains a Finite State Machine (FSM), regulates the SRs for the selection of BRAMs, with the chosen BRAM sending or receiving data to or from the APU.  The APP provides the solution to data synchronization through the BRAM banks. By managing the synchronization registers and the Sync controller, it ensures data consistency from different clock domains and guarantees that the APU processes data from the correct event, even with the presence of raw data input skew.

All trigger processing for a given Bunch Crossing (BC - an event in the detector) is handled in a single GEP. To process multiple events under significant throughput and latency constraints, the 48 GEP units operate in a round-robin fashion, where GEP1 processes data from BC1, followed by data from BC49, and so forth.
Data processing within the APUs of GEP is pipelined, such that upstream APUs may be processing data for BC49, while downstream APUs may still be processing data for BC1; in fact, we expect a plurality of BC's to be processed simultaneously within each GEP.

\subsection{\texttt{hls4ml}}
\begin{figure}[h!]
\centering\includegraphics[width=12cm]{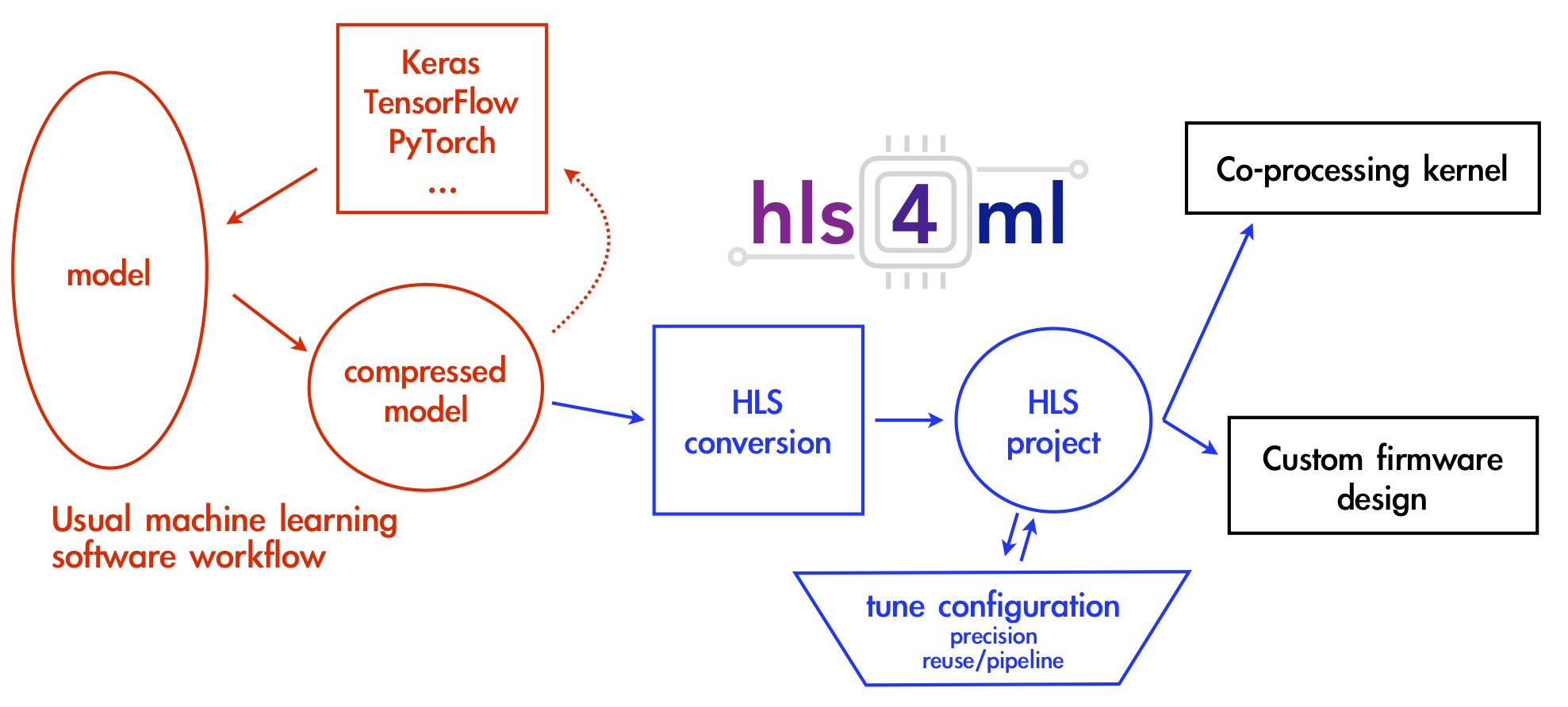}
\caption{\label{hls4ml_flow}The workflow of \texttt{hls4ml}, \texttt{hls4ml} will first read the model from Pytorch, Tensorflow, or Keras, then convert to the hardware descriptive language using Vivado HLS, and eventually to the FPGA}
\end{figure}

The trigger upgrade project aims to develop a low latency data processing system for high-energy physics. To help achieve this, the project is utilizing a high-level synthesis tool ~\cite{Duarte_2018} to convert machine learning models into FPGA firmware.  High-Level Synthesis for Machine Learning (\texttt{hls4ml}) is an open-source software package that provides a user-friendly interface for converting high-level machine learning models into hardware implementations. The tool generates hardware designs in hardware description languages (HDLs) such as VHDL or Verilog, which can then be synthesized and implemented on FPGAs. The workflow of \texttt{hls4ml} is: 1) automatically converting a machine learning model from TensorFlow~\cite{tensorflow2015-whitepaper}, Pytorch~\cite{NEURIPS2019_9015}, or Keras ~\cite{chollet2015keras} into an \texttt{hls4ml} project that is output in a hardware-oriented subset of C++; 2) using Vivado HLS to synthesize the C++ code into HDL; 3) Using Vivado to synthesize the HDL into an FPGA bitstream. Figure~\ref{hls4ml_flow} shows the workflow of \texttt{hls4ml}. \texttt{hls4ml} has been used in various high-energy physics experiments, including the Fermilab booster~\cite{StJohn:2020bpk}. 

\texttt{hls4ml} is a promising tool for APU designs for several reasons.  First, \texttt{hls4ml} is a convenient way to automatically convert a machine learning model into RTL, allowing for quick generation of different machine learning architectures. The user only needs to create the model using standard approaches in TensorFlow or Pytorch, and \texttt{hls4ml} can do the conversion to hardware. This saves designers significant amounts of time in implementing complex machine learning algorithms. Second, \texttt{hls4ml} can optimize hardware architectures for specific performance metrics, such as latency, throughput, or power consumption. This makes it a powerful tool for implementing real-time applications, such as those required by high-energy physics experiments. Third, \texttt{hls4ml} supports many different machine learning models, including dense neural network (DNN)~\cite{Duarte_2018}, convolution neural network (CNN)~\cite{Aarrestad:2021zos}, recurrent neural networks (RNN)~\cite{khoda2022ultralow}, and graph neural networks (GNNs)~\cite{Iiyama:2020wap, Heintz:2020soy}.

\subsection{\texttt{fwX}}
The software package \texttt{fwX} is used for implementing boosted decision tree(BDT)-based machine learning algorithms onto FPGAs for high-energy physics applications~\cite{Hong:2021snb,Carlson:2022vac, Roche:2023int}. Similar to \texttt{hls4ml}, it uses Vivado HLS to convert the model into RTL. It operates via a three-stage process: machine learning training with external software packages, optimization to fine-tune boosted BDT structures and parameters for physics performance and FPGA cost, and conversion to the firmware design through vendor tools.

The \texttt{fwX} software package has been used to implement nanosecond machine learning with deep decision trees that have been used for problems that include event classification, regression, and anomaly detection. These implementations have achieved high accuracy and low latency, making them suitable for real-time applications. The parallel decision paths architecture of \texttt{fwX} allows for efficient use of FPGA resources, resulting in high-performance implementations. Its ability to efficiently implement decision trees with large numbers of branches and leaves makes it a valuable tool for applications.
 
BDTs have been extensively utilized in high-energy physics applications, for instance in the discovery of the Higgs boson by the ATLAS and CMS collaborations~\cite{Aad:2012tfa, Chatrchyan:2012ufa}. In this context, \texttt{fwX} proves invaluable by efficiently implementing complex BDT models on FPGA, which has low latency (in nano second scale) and small resource usage.

The potential of \texttt{fwX} is underscored by its remarkable performance metrics. In one study ~\cite{Hong:2021snb}, for a complex BDT model with 100 training trees, a maximum depth of 4, and four input variables, it boasts a latency of only around 10 ns, or 3 clock ticks at 320 MHz. Notably, this level of performance is achieved with minimal resource utilization - less than 0.2\% of look-up tables and block RAM usage, less than 0.01\% of flip-flop usage, and no ultra RAM or digital signal processor (DSP) usage. This efficiency demonstrates \texttt{fwX}'s capacity to provide high-speed, low-resource implementations without compromising on the complexity or accuracy of the machine learning models.

\section{Experimental Result} 
\subsection{Deep Neural Network for B-tagging}
In the pursuit of refining particle identification within the ATLAS GEP, a DNN has been integrated into the APU, specifically focusing on a Jet tagging task. This task plays a crucial role in identifying the types of particles, particularly in distinguishing between different jet types, including those originating from b-quarks (B-tagging).

\begin{figure}[h]
\centering\includegraphics[width=8cm]{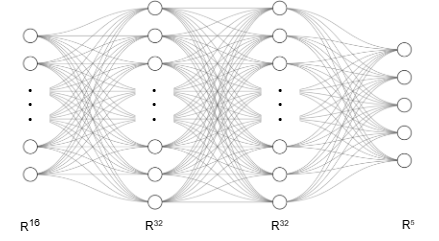}
\caption{The architecture of the dense neural network}
\label{dnn}
\end{figure}

The employed DNN model for B-tagging is structured with four dense layers consisting of 16, 32, 32, and 5 neurons, respectively. The final layer employs softmax activation for classifying input data into five distinct categories, tailored to differentiate various particle types accurately.  Figure~\ref{dnn} illustrates the DNN architecture, showcasing its layered structure and neuron configuration, which is pivotal for the B-tagging application.

The resource utilization of this DNN model is depicted in Table~\ref{DNN-resource}. The model demonstrates a balance between low latency and minimal resource usage, which is essential for real-time processing in the APU. With a latency of just 10 cycles, or 50ns at a 200MHz clock rate, this model exemplifies the feasibility of using \texttt{hls4ml}-generated machine learning models in APUs for high-energy physics experiments.

\begin{table}[]
\centering 
\caption{\label{DNN-resource}The resource usage of the B-tagging DNN model}
\begin{tabular}{|lll|}\hline  
Resource                & Utilization   & Utilization \% \\ \hline \hline 
DSP                     & 625           & 9.1   \\        
FF                      & 9646          & 0.41   \\     
LUT                     & 54441         & 4.6   \\      
BRAM                    & 18            & 0.83   \\      
\hline  
\end{tabular}
\end{table}

This B-tagging DNN model not only fulfills the real-time processing requirements but also highlights the effectiveness of implementing advanced machine learning techniques in the field of high-energy physics. The efficient use of FPGA resources, combined with the high-speed processing capabilities, positions this approach as a valuable asset for current and future experiments in the ATLAS GEP.

\subsection{VBF Classification in BDTs}
Machine learning algorithms in the form of neural networks and BDTs are commonly used to separate signals and backgrounds in high-energy physics experiments. Examples include hadronic $\tau$ lepton identification~\cite{CMS:2022prd} and identification of jets that contain a $b$-hadron~\cite{ATLAS:2018sgt}.

As an example for BDT classification in the ATLAS GEP, we use the problem of separating vector boson fusion Higgs production from multijet background. 
We utilize the samples produced for the \texttt{fwX} classification paper~\cite{Hong:2021snb}. 

Further details, as well as input distributions, are available in the \texttt{fwX} paper~\cite{Hong:2021snb} and the corresponding public dataset~\cite{Roche:classifier}. 
As the VBF trigger is dominated by high transverse momentum ($p_{T}$) jets, we assume that the hardware studies performed will be a reasonable representation of the GEP performance. 

The classifier is trained using kinematic variables corresponding to the two VBF jets. These include the transverse momentum of the sub-leading jet $p_{T2}$, and calculated quantities on the two VBF jets. These calculated quantities include the vector sum  $p_{T}(jj)$, the scalar sum, $H_{T}$(jj), and the invariant mass of the two jets $m_{jj}$. To account for jets in opposite hemispheres of the detector, the product of the two jet pseudo-rapidity values are computed: $\eta_{1}\cdot\eta_{2}$. The range and number of bits assigned to each input variable is summarized in Table~\ref{VBF-in}. 

\begin{table}[]
\centering 
\caption{\label{VBF-in}Input variables, range of each variable and number of bit assigned to each variable.}
\begin{tabular}{|lll|}\hline  
Variable                & Range         & bits \\ \hline \hline 
$\eta_{1}\cdot\eta_{2}$ & -20--20       & 12   \\
$p_{T2}$               & 0 -- 1000 GeV & 12   \\
$p_{T}$(jj)             & 0 -- 1500 GeV & 12   \\
$H_{T}$(jj)             & 0 -- 1500 GeV & 12  \\
$m_{jj}$                & 0 -- 4500 GeV & 7    \\
\hline  
\end{tabular}
\end{table}

The BDT model is trained using the TMVA~\cite{Hocker:2007ht} package, which implements the AdaBoost~\cite{adaboost} method with 100 trees and a max depth of 4. During the simplification step performed by \texttt{fwX}, the number of trees was reduced to 10. 

The performance of the model implemented in the APU is evaluated by examining the latency, as well as the FPGA resource costs using the Xilinx FPGA VU9P chip. The latency was evaluated to be 7 clock cycles with the clock running at a rate of 320MHz, which means the latency is 21.875ns. The resource usage is shown in Table~\ref{BDT-resource}. These results underscore the extremely low resource consumption on the FPGA, showcasing its practicality and effectiveness.

\begin{table}[]
\centering 
\caption{\label{BDT-resource}The resource usage of the classification BDT model}
\begin{tabular}{|lll|}\hline  
Resource                & Utilization           & Utilization \% \\ \hline \hline 
DSP                     & 2             & 0.029  \\
FF                      & 597           & 0.025   \\
LUT                     & 2756          & 0.23   \\
BRAM                    & 48            & 2.2   \\
\hline  
\end{tabular}
\end{table}

\subsection{Missing Transverse Momentum Regression BDT}
{
Regression models are useful for a wide variety of physics applications, including reconstruction of missing transverse momentum, $E_{T}^{\text{miss}}$~\cite{CMS:2019ctu} and hadronic $\tau$ leptons~\cite{ATLAS:2015boj}. 
To evaluate the hardware performance of a regression model in the APU, 
a regression model to evaluate $E_{T}^{\text{miss}}$ is studied. 
The implementation in the \texttt{fwX} regression studies was originally performed using public Delphes samples~\cite{Carlson:regression} described in Ref~\cite{Carlson:2022vac}.

In particular, this model is trained to identify the true $E_{T}^{\text{miss}}$ based on a simulated sample of Higgs boson events that decay to neutrinos that do not interact with the detector. The eight input variables are described in Ref~\cite{Carlson:2022vac}. The regression model is configured with 40 trees, a tree depth of 6. 

The performance of the model implemented in the APU is evaluated by examining the latency, as well as the FPGA resource costs using the Xilinx FPGA VU9P chip. The latency was evaluated to be 11 clock cycles with the clock running at a rate of 320MHz, which makes the latency 34 nanoseconds. The resource usage is shown in the Table ~\ref{regression-resource}.

\begin{table}[]
\centering 
\caption{\label{regression-resource}The resource usage of the regression BDT model, post-synthesis. The utilization is given in the total number of available units utilized as well as the fraction available on the FPGA in \%.}
\begin{tabular}{|lll|}\hline  
Resource                & Utilization           & Utilization (\%) \\ \hline \hline 
DSP                     & 0         & 0.0   \\
FF                      & 1987      & 0.084   \\
LUT                     & 3493     & 0.30  \\
BRAM                    & 12        & 0.56   \\
\hline  
\end{tabular}
\end{table}
}
\subsection{Quark-Gluon Jet Tagging Algorithm} 
The capacity to distinguish between quark-originated and gluon-originated jets is widely applicable to numerous physics investigations at the LHC~\cite{ATL-PHYS-PUB-2017-017,Lee_2019,Komiske_2017}. This section introduces a technique for differentiating quark-based and gluon-based jets by employing a deep neural network classifier that analyzes the complete radiation pattern within a jet as an image. The energy deposits in the calorimeters serve as inputs for the jet reconstruction and classification algorithm. The energy deposit organization scheme makes use of topological calorimeter-cell clusters (topo-clusters)~\cite{aaboud2017jet}. Topo-clusters are used as input for jet reconstruction with the anti-$k_t$ jet algorithm~\cite{cacciari2008anti} with distance parameter $R = 0.4$. 
Jets labeled as gluon or quark (excluding top quark) are considered. Jets with transverse momentum ($p_T$) between 50 and 75 GeV and $|\eta| < 2.5$ are selected where $\eta$ is the pseudorapidity. Jets are required to satisfy generator-level matching criteria: the jet must be matched to a parton-level quark or gluon and all of its decay products within $\Delta R = 0.4$ where $\Delta R = \sqrt{(\Delta\eta)^2 + (\Delta\phi)^2}$ and $\phi$ is the azimuthal angle.

As a first step in constructing a jet image, the constituents inside a jet are translated in $\eta$ and $\phi$ so that the jet's center is located at the center in $\eta$-$\phi$ space. Then, a fixed grid of size $15 \times 15$ in $\eta$ and $\phi$ with pixel sizes $0.055 \times 0.055$ is centered on the origin. The intensity of each pixel is the total $E_T$ within the pixel, using topocluster input. Pixel values are then normalized by dividing them by the value of the hottest (maximum) pixel in the image. This scaling ensures that the pixel values of the entire image are between 0 and 1. Then, the pixel values are scaled to a range between 0 and 255, this is done by multiplying each pixel value by 255.

In this study, we utilize images of jets as input for a deep neural network classifier, specifically a deep CNN. The CNN~\cite{Goodfellow-et-al-2016} architecture we employ involves a convolutional layer with ReLU activation, paired with a Max-pooling layer. The network outputs a softmax function that predicts the probability of a quark or gluon jet. The convolutional layer includes 4 filters with filter sizes of 2x2, while the Max-pooling layers perform a 2x2 down sampling. To avoid overfitting, we employ dropout on the convolutional and final fully connected layers at a rate of 0.1. Training is performed by minimizing the binary cross-entropy, using the Adam optimizer~\cite{kingma2017adam} implemented in Keras with a learning rate of 0.0001 over 100 iterations and a batch size of 256. The training dataset contains approximately 105K events, while the test dataset consists of around 26K events.

For this CNN algorithm, we convert it into an FPGA implementation via the \texttt{hls4ml} toolchain. The performance of the model implemented in the APU is evaluated by examining the latency, as well as the FPGA resource costs using the Xilinx FPGA VU9P chip. The latency was evaluated to be 233 clock cycles with the clock running at a rate of 200MHz, which makes the latency 1.2 microseconds. The resource usage is shown in Table~\ref{qg_tagger-resource}.

\begin{table}[]
\centering 
\caption{\label{qg_tagger-resource}The resource usage of the qg tagger CNN model}
\begin{tabular}{|lll|}\hline  
Resource                & Utilization           & Utilization \% \\ \hline \hline 
DSP                     & 305                  & 4.5   \\
FF                      & 4812                 & 0.20   \\
LUT                     & 7504                 & 0.63   \\
BRAM                    & 9                    & 0.42   \\
\hline  
\end{tabular}
\end{table}

\section{Conclusion}
In this paper, we developed mechanisms to easily implement machine learning based algorithms into the Algorithm Processing Unit for the ATLAS Global Event Processor.  We tested BDT and Neural Network models prepared using the \texttt{fwX} and \texttt{hls4ml} tools respectively. 

Our study underscores the efficacy of machine learning tools when integrated into the APU framework, as demonstrated by the performance evaluation presented in Table \ref{tab:model_complexities}. The various machine learning models, ranging from the VBF classifier to the more complex q/g CNN, are implemented with impressive efficiency, maintaining latency values from as low as 22ns up to 1.2$\mu s$. Notably, the resource utilization for these models remains commendably low, with less than 10\% of the total resources of the FPGA VCU118 being employed, even for the more resource-intensive B-tagging DNN. This data indicates not only the high efficiency of our integrated ML models but also showcases the scalable complexity of the models that the APU can support. The proportional increase in resource usage, such as the LUT and DSP consumption, aligns with the enhanced capabilities and complexities of the respective algorithms, thereby validating the APU's capability to execute advanced computational tasks within the stringent requirements set by the GEP. 

As we look to the future, this work lays the groundwork for the integration of increasingly complex machine learning models, which could further enhance the performance of APU. The methodologies presented in this paper have potential applications in various experimental setups, thereby contributing to the continuous improvement and evolution of real-time data processing systems. With ongoing advancements in machine learning and FPGA technologies, the application of tools such as \texttt{hls4ml} and \texttt{fwX} may become even more critical at the nexus of high-energy physics and real-time data processing. For instance, the deployment of RNN implementations on FPGAs, as discussed in ~\cite{khoda2022ultralow}, or the advancements in real-time data processing illustrated in ~\cite{10190588,ghielmetti2022realtime}, exemplify the expanding scope of these technologies.

Overall, this work emphasizes the ability to easily deploy the \texttt{hls4ml} and \texttt{fwX} tools, demonstrating their successful application in meeting the needs of the next generation of the LHC's high-speed data processing systems.

\begin{table}[h!]
\centering
\caption{Comparison of Model Complexities}
\label{tab:model_complexities}
\begin{tabular}{|l|c|c|c|c|}
\hline
                     & \textbf{VBF classifier} & \textbf{MET regression} & \textbf{B-tagging DNN}   & \textbf{q/g CNN} \\ \hline
\textbf{Tool}        & \texttt{fwX} (Depth = 4)         & \texttt{fwX} (Depth = 6)         & \texttt{hls4ml}                   & \texttt{hls4ml}           \\ \hline
\textbf{Clock}       & 320 MHz                 & 320 MHz                 & 200 MHz                  & 200 MHz          \\ \hline
\textbf{Latency}     & 22 ns                   & 34 ns                   & 50 ns                    & 1.2 us           \\ \hline
\textbf{LUT}         & 0.23\%                  & 0.30\%                  & 4.6\%                    & 0.63\%           \\ \hline
\textbf{DSP}         & 0.029\%                 & 0.0\%                   & 9.1\%                    & 4.5\%            \\ \hline
\textbf{FF}          & 0.025\%                 & 0.084\%                 & 0.41\%                   & 0.20\%           \\ \hline
\textbf{BRAM}        & 2.2\%                   & 0.56\%                  & 0.83\%                   & 0.42\%           \\ \hline
\end{tabular}
\end{table}

\section*{Acknowledgments}
We acknowledge the ATLAS Global Even Processor group as a supportive community of experts and collaborators. This group was important for the development of this project. We particularly thanks Wade Fisher offers clear guidance to conduct this project.

Jiang, Hauck and Hsu are supported by National Science Foundation (NSF) grants No. 2117997. 
Carlson is supported by NSF grant No. 2209370 and No. 2117997 and would like to thank Steve Roche for technical support with \texttt{fwX}. 

\bibliographystyle{JHEP}
\bibliography{biblio.bib}

\end{document}